\documentstyle[12pt]{article}
\parskip=\medskipamount

\title{A covariant action for the eleven dimensional superstring}
\author{A.A. Deriglazov and A.V.
Galajinsky\thanks{deriglaz@phys.tsu.tomsk.su; \quad galajin@phys.tsu.tomsk.su\/}}
\date{Department of Mathematical Physics,\\
Tomsk Polytechnical University, 634004 Tomsk, Russia}
\begin{document}
\maketitle
\large
\begin{abstract}
We suggest a super Poincar\'e invariant action for closed eleven
dimensional superstring. The sector of physical variables  $x^i$, $\theta_a$,
$\bar\theta_{\dot a}$, with $a,\dot a=1\dots8$ and $x^i$ the transverse
part of the $D=11$ $x^\mu$ coordinate is shown to possess free dynamics.
\end{abstract}

\section{Introduction}
The classical Green--Schwarz (GS) superstring (with the manifest
space-time supersymmetry and local $\kappa$-symmetry) can move in
spacetime dimensions 3, 4, 6, and 10 [1]. Thus, the standard approach
fails to construct a $D=11$ superstring action, while known results (see,
for example, [2--4] and references therein) suggest the existence of a
consistent quantum theory incorporating the $D=11$ supergravity. At present
moment, the latter can be viewed as either strong coupling limit of the
type IIA superstring [5], or as an effective theory of the supermembrane [6],
or it may be regarded as a constituent of $M$-theory [2, 5, 7]. The
purpose of this letter is to present some results in this direction.

The crucial ingredient in construction of GS superstring action is the
$\Gamma$-matrix identity
\begin{equation}
\Gamma^\mu_{\alpha(\beta}(C\Gamma^\mu)_{\gamma\delta)}=0.
\end{equation}
which provides both global supersymmetry and local $\kappa$-symmetry for
the superstring action [8]. The $\kappa$-symmetry, in its
turn, ensures free dynamics in the sector of physical variables. To
elucidate construction that will be suggested below for $D=11$,
let us discuss the problem in the Hamiltonian framework, where
one faces with the well known fermionic constraints (see, for example,
Refs. 1 and 8)
\begin{equation}
L_\alpha\equiv p_{\theta\alpha}-i(\bar\theta\Gamma^\mu)_\alpha(p_\mu
+\Pi_{1\mu})=0,
\end{equation}
obeying the Poisson bracket
\begin{equation}
\{ L_\alpha,L_\beta\}=2i(\hat p^\mu+\Pi^\mu_1)\Gamma^\mu_{\alpha\beta}
\delta(\sigma-\sigma')-2\bar\theta^\gamma\partial_1\theta^\delta
\Gamma^\mu_{\gamma(\delta}(C\Gamma^\mu)_{\alpha\beta)}
\delta(\sigma-\sigma').
\end{equation}
By virtue of Eq. (1) the last term in Eq. (3) vanishes for
$D=3,4,6,10$. The resulting equation then means that half of the
constraints are first class. This exactly corresponds to the local
$\kappa$-symmetry presented in the Lagrangian framework.

The next step is to impose an appropriate gauge. Then the full system
(constraints and gauges) looks as follows:
\begin{eqnarray}
&& L_\alpha=0,\\
&& \Gamma^+\theta=0,
\end{eqnarray}
and is second class (without making use of Eq. (1) ).

The situation changes drastically for the $D=11$ case, where instead of
Eq. (1), one has [9, 10]
\begin{equation}
10 \Gamma^\mu_{\alpha(\beta}(C\Gamma^\mu)_{\gamma\delta)}+
\Gamma^{\mu\nu}_{\alpha(\beta}(C\Gamma^{\mu\nu})_{\gamma\delta)}=0.
\end{equation}
Being appropriate for the construction of the supermembrane action [6],
this identity does not allow one to formulate a $D=11$ superstring with
desirable properties. As was shown by Curtright [9], the globally
supersymmetric action based on this identity involves additional to
$x^i$, $\theta_a$, $\bar\theta_{\dot a}$ degrees of freedom in the
physical sector. Moreover, it does not possess $\kappa$-symmetry
which might provide free dynamics [9, 10].

In this letter we suggest a $D=11$ super Poincar\'e invariant action
for the classical closed superstring which possesses free dynamics in the
physical variables sector.
Instead of the standard approach which implies the search for an action
with a local $\kappa$-symmetry (or, equivalently, with the corresponding
first class constraints), we present a theory in which constraints
like Eqs. (4), (5) arise among others. Since they are second class,
the $\kappa$-symmetry and the identity (6) are not
necessary for the construction. Thus, at the classical level,
a superstring of the type described can exist in any spacetime dimensions
and the known brane scan [4] can be revised. In particular,being
applied to the $D=10$ case, our construction yield the model in which
spectrum of physical states coincide with those of
$N=1$, $D=10$ Green--Schwarz superstring. For definiteness, in this letter
we discuss the $D=11$ case only.

Two comments are in order. First, one needs to covariantize Eq. (5),
and the simplest possibility is $\Lambda_\mu\Gamma^\mu\theta=0$, with
$\Lambda^2=0$. It is assumed that the additional vector variable
$\Lambda^\mu$ is introduced in such a way that the gauge $\Lambda^-=1$
is possible. Second, one can expect that a model with constraints like
Eqs. (4) and (5) will possess (if any) off-shell super Poincar\'e symmetry
in a nonstandard realization. Actually, global supersymmetry which does
not spoil the equation $\Lambda_\mu\Gamma^\mu\theta=0$ turns out to be
$\delta\theta\sim\Lambda_\mu\Gamma^\mu\epsilon$. On-shell, where
$\Lambda^2=0$, only half of the supersymmetry parameters
$\epsilon^\alpha$ are essential.

It is worth mentioning other motivation for this work.
The action for the super $D$-brane which allows local $\kappa$-symmetry
is rather complicated [11, 15]. One can hope, that being applied to that
case, our method will lead to a more simple formulation.

The work is organized as follows. In Sect. 2 the action and its local
symmetries are presented. In Sec. 3 within the framework of the Hamiltonian
approach we prove that the model possesses free dynamics. In Sect. 4 an
off-shell realization of the super Poincar\'e algebra is obtained
and discussed.

{\it Notations}. We use $32\times 32$ $\Gamma$-matrices in the Majorana
representation [16]. They have the properties $(\Gamma^0)^T=-\Gamma^0$,
$(\Gamma^i)^T=\Gamma^i$, $(\Gamma^\mu)^*=\Gamma^\mu$ and obey the
algebra $\{\Gamma^\mu,\Gamma^\nu\}=-2\eta^{\mu\nu}$, $\eta^{\mu\nu}=
(+,-,\dots,-)$. Charge conjugate of the Majorana spinor $\theta^\alpha=
(\theta_a,\bar\theta'_{\dot a},\theta'_a,\bar\theta_{\dot a})$,
$\alpha=1,2,\dots,32$, $a,\dot a=1\dots8$ is $\bar\theta=\theta C$ with
$C=\Gamma^0$. It will be convenient to use the following light-cone
$\Gamma$-matrices:
\[ \Gamma^\pm=\frac 1{\sqrt 2}(\Gamma^0\pm\Gamma^9), \quad \Gamma^i, \quad
i=1,2,\dots,8,10. \]
Momenta, conjugate to the configuration space variables $q^i$ are
denoted as $p_{qi}$.

\section{Action and its local symmetries}

The action functional to be examined is
\begin{eqnarray}
S=\int d^2\sigma\Big\{\frac{-g^{ab}}{2\sqrt{-g}}\Pi_a^\mu\Pi^\mu_b-
i\varepsilon^{ab}\partial_a x^\mu(\bar\theta\Gamma^\mu\partial_b\theta)-\cr
-i\Lambda^\mu\bar\psi\Gamma^\mu\theta-\frac 1\phi \Lambda^\mu
\Lambda^\mu -\Lambda^\mu\varepsilon^{ab}\partial_a A^\mu_b\Big\},
\end{eqnarray}
where it was denoted $\Pi^\mu_a\equiv \partial_a x^\mu -i\bar\theta
\Gamma^\mu \partial_a\theta$. The first two terms are exactly those of
the GS superstring action written in eleven dimensions.
The origin of the remaining terms is as follows. The third and the
fourth terms will supply the appearance of the equations
$\Lambda_\mu\Gamma_\mu\theta=0$ and $\Lambda^2=0$. Hence, the variables
$\bar\psi^\alpha$ and $\phi$ can be viewed as the Lagrange multipliers
enforsing the constraints.
The last term was added to suppress the appearance of some undesirable
constraints in addition to those mentioned above. The expression of such
a kind was successfully used before [17, 18] in a different context.

Note also that the Wess--Zumino term in the $D=10$ GS action provides
the appearance of the local $\kappa$-symmetry [1]. In our model it plays
a different role, as shown below.

Let us briefly comment on the structure of local symmetries
for the action (7).
Local bosonic symmetries include $d=2$ reparametrizations, Weyl symmetry,
and the following transformations with parameters $\xi^\mu(\sigma^a)$
and $\omega_a(\sigma^b)$:
\begin{equation}
\delta A^\mu_a=\partial_a\xi^\mu+\omega_a\Lambda^\mu, \qquad
\delta\phi=\frac 12 \phi^2\varepsilon^{ab}\partial_a \omega_b.
\end{equation}
These symmetries are reducible because their combination with
parameters of special form $\omega_a=\partial_a\omega$, $\xi^\mu=
-\omega\Lambda^\mu$ is a trivial symmetry: $\delta A^\mu_a=-\omega
\partial_a\Lambda^\mu$, $\delta\phi=0$ (note that
$\partial_a\Lambda^\mu=0$ is one of the equations of motion). Thus, Eq.
(8) includes 12 essential parameters which correspond to the primary first
class constraints $p^\mu_0\approx 0$, $p_\phi\approx 0$ (see below).

There is also a fermionic symmetry with the parameters
$\xi^\alpha(\sigma^a)$:
\begin{equation}
\delta\bar\psi=\bar\xi\Gamma^\mu\Lambda_\mu, \qquad
\delta\phi=-\phi^2(\bar\xi\theta),
\end{equation}
from which only 16 are essential on-shell where $\Lambda^2=0$. As shown
below, reducibility of this symmetry make no special problem for
covariant quantization.

Let us present arguments that the action constructed describes a free
theory. Equations of motion for the action (7) are
$$
\Pi^\mu_a\Pi^\mu_b-\frac 12 g_{ab}(g^{cd}\Pi^\mu_c\Pi^\mu_d)=0,
\eqno{(10.a)}$$
$$
\partial_a\left(\frac{g^{ab}}{\sqrt{-g}}\Pi^\mu_b+i\varepsilon^{ab}
\bar\theta\Gamma^\mu\partial_b\theta\right)=0,
\eqno{(10.b)}$$
$$
4i\Pi^\mu_b(\Gamma^\mu P^{-ba}\partial_a\theta)_\alpha+
\varepsilon^{ab}\theta^\beta\partial_a\theta^\gamma\partial_b
\theta^\delta\Gamma^\mu_{\alpha(\beta}C\Gamma^\mu_{\gamma\delta)}+
i\Lambda^\mu(\Gamma^\mu\psi)_\alpha=0,
\eqno{(10.c)}$$
$$
\Lambda^\mu\Gamma^\mu\theta=0, \qquad \Lambda^2=0,
\eqno{(10.d)}$$
$$
\partial_a\Lambda^\mu=0, \qquad \varepsilon^{ab}\partial_a A^\mu_b+
\frac 2\phi\Lambda^\mu+i\bar\psi\Gamma^\mu\theta=0,
\eqno{(10.e)}$$
\addtocounter{equation}{1}
where
\[ P^{-ba}=\frac 12\left(\frac{g^{ba}}{\sqrt{-g}}-\varepsilon^{ba}
\right). \]
Multiplying Eq. (10.c) with $\Lambda_\mu\Gamma^\mu$ one gets
\begin{equation}
(\Lambda^\mu\Pi^\mu_b)P^{-ba}\partial_a\theta=0.
\end{equation}
In the gauge $\Lambda^-=1$, supplemented by the conformal one, this can
be rewritten as
\begin{equation}
(\partial_0+\partial_1)\theta=0.
\end{equation}
Hence, any solution $\theta(\sigma)$ of the system
(10) obeys the free equation, which is accompanied with
$\Lambda_\mu\Gamma^\mu\theta=0$. The latter reduces to
$\Gamma^+\theta=0$ in the gauge chosen.
Thus, Eqs. (10.a--c) for the $g^{ab}$, $x^\mu$, $\theta^\alpha$ variables
look like those of the GS superstring. In the result one expects free
dynamics in this sector provided that the conformal gauge has been assumed.
In the next section we will rigorously prove this fact by direct
calculation in the Hamiltonian framework.

\section{Analysis of dynamics}

From the explicit form of the action functional it follows that the
variable $\Lambda^\mu$ can be excluded by making use of its equation of
motion. The Hamiltonian analog of the situation is a pair of second
class constraints ${p_\Lambda}^\mu=0$, $p_{A_1}{}^\mu-\Lambda^\mu=0$,
which can be omitted after introducing the associated Dirac bracket.
The Dirac brackets for the remaining variables prove to coincide with
the Poisson ones. The Hamiltonian looks like
\begin{eqnarray}
& H=\displaystyle\int d\sigma\Big\{-\frac N2(\hat p^2+\Pi_{1\mu}\Pi_1^\mu)-N_1
\hat p_\mu \Pi_1^\mu +p_{1\mu}(\partial_1 A_0^\mu+i\bar\psi\Gamma^\mu\theta)+ \cr
& +\displaystyle\frac 1\phi (p_1^\mu)^2+ \lambda_\phi\pi_\phi +
\lambda_{0\mu}p_0^\mu +\lambda^{ab}(\pi_g)_{ab}+{\lambda_\psi}^\alpha
p_{\psi\alpha} +L_\alpha{\lambda_\theta}^\alpha\Big\},
\end{eqnarray}
where $p^\mu,p_0^\mu,p_1^\mu$ are momenta conjugate to the variables
$x^\mu, A_0^\mu,A_1^\mu$, respectively, and $\lambda_*$ are the Lagrange
multipliers corresponding to the primary constraints. In Eq. (13) we
also denoted
\begin{eqnarray*}
& N = \displaystyle\frac{\sqrt{-g}}{g^{00}}, \quad N_1=\frac{g^{01}}{g^{00}}, \quad
\hat p^\mu=p^\mu-i\bar\theta\Gamma^\mu \partial_1\theta, \\
& L_\alpha\equiv p_{\theta\alpha}-i(p^\mu+\Pi_1^\mu)(\bar\theta
\Gamma^\mu)_\alpha=0.
\end{eqnarray*}
Detailed analysis shows that the constraints $(\pi_g)_{ab}=0$ are first
class\footnote{Note that the constraints $(\pi_g)_{ab}=0$ are not
separated from the $S_\alpha$ in Eq. (12). An appropriate modification
is
\[ (\tilde\pi_g)_{ab}\equiv (\pi_g)_{ab}+\frac 1{2(\hat p+\Pi_1)p_1}
(p_\psi\Gamma^\mu \Gamma^\nu\theta)(\hat p^\mu+\Pi_1^\mu){T^v}_{ab}, \]
where the coefficients $T^v_{ab}$ can be extracted from the equality
$\{(\pi_g)_{ab},S_\alpha\}={T^\mu}_{ab}(\bar\theta\Gamma^\mu)_\alpha$.},
which suggests the gauge choice $g^{ab}=\eta^{ab}$. Then the full set of
constraints can be written in the form
\begin{eqnarray}
&& \pi_\phi=0, \qquad p_0^\mu=0;\\
&& (p_1^\mu)^2=0, \qquad \partial_1p_1^\mu=0, \qquad (\hat p^\mu \pm
\Pi_1^\mu)^2=0,\cr
&& L_\alpha=0, \qquad \bar\theta\Gamma^\mu p_{1\mu}=0,\\
&& p_{\psi\alpha}=0, \qquad S_\alpha\equiv \bar\psi \Gamma^\mu p_{1\mu}
+ (\bar\theta \Gamma^\mu)_\alpha D_\mu=0.\nonumber
\end{eqnarray}
where
\begin{equation}
D^\mu\equiv \xi(\hat p^\mu+\Pi_1^\mu)-\partial_1 p^\mu, \qquad
\xi\equiv \frac{\partial_1 \hat p^\mu p_{1\mu}}{(\hat p^\nu +\Pi_1^\nu)
p_{1\nu}}.
\end{equation}
Besides, some of the Lagrange multipliers have been determined in the
process
\begin{equation}
\bar\lambda_\theta=\partial_1\bar\theta +\frac \xi 2 \bar\theta, \qquad
\lambda_1^\mu=\partial_1 A_0^\mu +\frac 2\phi A_1^\mu +i\bar\psi
\Gamma^\mu\theta.
\end{equation}
It is interesting to note that the constraints $S_\alpha=0$ appear as
tertiary ones in the Dirac--Bergmann algorithm. It is worth
mentioning also that the fermionic constraints $L_\alpha=0$ obey the algebra
(3), and being considered on their own (without making use the
constraints $\bar\theta \Gamma^\mu p_{1\mu}=0$) form a system which
has no definite class (it corresponds to the absence of
the $\kappa$-symmetry in the GS action written in eleven dimensions).

To go further, let us impose gauge fixing conditions to the first class
constraints (14). The choice consistent with the equations of motion is
\begin{equation}
\phi=2, \qquad A_0^\mu=-i\int_0^\sigma d\sigma' \bar\psi \Gamma^\mu
\theta.
\end{equation}
After that, dynamics for the remaining variables is governed by
$$
\begin{array}{ll} \partial_0\psi^\alpha={\lambda_\psi}^\alpha,
\qquad & \partial_0 p_{\psi\alpha}=0,\\
p_{\psi\alpha}=0, & S_\alpha=0;\end{array}
\eqno{(19.a)}$$
$$
\begin{array}{ll} \partial_0 A_1^\mu=p_1^\mu,
\qquad & \partial_0 p_1^\mu=0,\\
(p_1^\mu)^2=0, & \partial_1 p_1^\mu=0;\end{array}
\eqno{(19.b)}$$
$$
\begin{array}{c} \partial_0 x^\mu=-p^\mu,
\qquad \partial_0 p^\mu=-\partial_1\partial_1 x^\mu,\\
(\hat p^\mu\pm\Pi_1^\mu)^2=0;\end{array}
\eqno{(19.c)}$$
$$
\begin{array}{c} \partial_0\theta=-\partial_1\theta-\displaystyle\frac \xi 2 \theta,\\
L_\alpha=0, \qquad (\bar\theta \Gamma^\mu)_\alpha p_{1\mu}=0.\end{array}
\eqno{(19.d)}$$
\addtocounter{equation}{1}
The sector (19.a) includes $ 32+16$ independent constraints from which
the first class ones can be picked out as follows:
\begin{equation}
(p_\psi \Gamma^\mu)_\alpha p_{1\mu}=0.
\end{equation}
Let us impose the following covariant (and redundant) gauge fixing
conditions to Eq. (20)
\begin{equation}
{S^1}_\alpha \equiv \frac 1{(\hat p+\Pi_1)p_1} \bar\psi \Gamma^\mu
(\hat p_\mu+\Pi_{1\mu})=0.
\end{equation}
Then the set of equations $S_\alpha=0$, ${S^1}_\alpha=0$ is equivalent
to
\begin{equation}
\tilde S \equiv \bar\psi -\frac 1{2(\hat p+\Pi_1)p_1} \bar\theta
\Gamma^\mu D_\mu \Gamma^\nu (\hat p_\nu +\Pi_{1\nu}),
\end{equation}
which forms a nondegenerate Poisson bracket with the
constraint $p_{\psi\alpha}=0$
\begin{equation}
\{ p_{\psi\alpha}, \tilde S_\beta\}=-C_{\alpha\beta}.
\end{equation}
After transition to the Dirac bracket associated with the second class
functions $p_{\psi\alpha}$, $\tilde S_\alpha$, the variables $\psi$,
$p_\psi$ can be dropped.

To get dynamics in the final form, we pass to the light-cone coordinates
$x^\mu \to (x^+,x^-,x^i)$, $i=1,2,\dots,8,10$, $\theta^\alpha \to
(\theta_a, \bar\theta'_{\dot a}, \theta''_a, \bar\theta_{\dot a})$,
$a,\dot a=1,\dots,8$ and impose the gauge fixing conditions
\begin{equation}
A^-_1=\tau, \qquad A^+_1=A_1^i=0, \qquad x^+=P^+\tau, \qquad
p^+=-P^+={\rm const}
\end{equation}
to the remaining first class constraints from Eqs. (19..b), (19.c). Then it
is easy to show that $32+16$ constraints $L_\alpha=0$, $\bar\theta\Gamma^\mu
p_{1\mu}=0$ are second class. One gets also that $p_1^-=1$, $p_1^+=p_1^i=0$,
while the equation $\bar\theta\Gamma^\mu p_{1\mu}=0$ acquires the form
$\Gamma^+\theta=0$. The solution is $\theta^\alpha=(\theta_a,0,0,\bar
\theta_{\dot a})$, with $\theta_a$ and $\bar\theta_{\dot a}$ the
$SO(8)$ spinors of opposite chirality. Note that the condition $p^-_1=1$
is consistent with the closed string boundary conditions only.
It is worth mentioning also that in the
gauge chosen the relation $(\hat p^\mu+\Pi_1^\mu)p_{1\mu}\ne0$ holds,
which correlates with the assumption made above in Eqs. (16), (21). For
the remaining variables one gets free equations
\begin{equation}
\begin{array}{cc} \partial_0 x^i=-p^i, \qquad &
\partial_0 p^i=-\partial_1 \partial_1 x^i;\\
(\partial_0+\partial_1)\theta_a=0, &
(\partial_0+\partial_1)\bar\theta_{\dot a}=0\end{array}
\end{equation}
which look similar to those of $D=10$ GS superstring.
Moreover, $\theta_a$ and $\bar\theta_{\dot a}$ form two pairs of
selfconjugate variables under the Dirac bracket, associated with the
constraints (15)
\begin{equation}
\{\theta_a,\theta_b\}=\frac i{\sqrt 8 P^+}\delta_{ab}, \qquad
\{\bar\theta_{\dot a},\bar\theta_{\dot b}\}=\frac i{\sqrt 8 P^+}
\delta_{\dot a\dot b}.
\end{equation}

It is interesting to note that omitting the Wess--Zumino term in Eq.
(7) one arrives at the theory which possesses all the properties of the
model (7) with the only modification in Eq. (25): $(\partial_0-
c\partial_1)\theta=0$, with $c$ a constant. Depending on the gauge
chosen it can take any value except $c=\pm1$. So the dynamics is not
manifestly $d=2$ Poincar\'e covariant, provided that $\theta$ is a
$d=2$ scalar. It is the Wess--Zumino term which corrects this
inconsistency.

\section{Off-shell realization of the $D=11$ super-Poincar\'e algebra}

It is convenient first to recall the situation for $D=10$ GS
superstring. Off-shell realization of the super Poincar\'e algebra for
that case includes the Poincar\'e transformations accompanied with the
supersymmetries
\begin{equation}
\delta\theta^\alpha=\epsilon^\alpha, \qquad
\delta x^\mu=-i\bar\theta\Gamma^\mu\epsilon.
\end{equation}
Being considered on their own, these transformations in the gauge
$\Gamma^+\theta=0$ are reduced to trivial shifts in the sector of physical
variables
\begin{equation}
\delta\bar\theta_{\dot a}=\bar\epsilon_{\dot a}, \qquad
\delta x^i=0.
\end{equation}
To get on-shell realization of the supersymmetry algebra, one needs to
consider a combination of the $\epsilon$- and $\kappa$-transformations
$\delta_\epsilon+\delta_{\kappa(\epsilon)}$, which does not spoil the
gauge $\Gamma^+\theta=0$. These transformations are (see, for example,
Ref. 19)
\begin{equation}
\delta\bar\theta_{\dot a}=\bar\epsilon_{\dot a}+\frac 1{P^+}
\partial_- x^i \bar\gamma^i{}_{\dot aa}\epsilon_a, \qquad
\delta x^i=-i\sqrt{2}(\bar\theta\bar\gamma^i\epsilon).
\end{equation}

We turn now to the $D=11$ case. Off-shell realization of the super
Poincar\'e algebra for the action (7) includes the Poincar\'e
transformations in the standard realization and the following
supersymmetries with 32-component spinor parameter $\epsilon^\alpha$
\begin{eqnarray}
&& \delta\theta=\tilde\Lambda\epsilon, \qquad
\delta x^\mu=-i\bar\theta\Gamma^\mu\tilde\Lambda\epsilon,\cr
&& \delta {A^\mu}_a=-2i\epsilon_{ab}\displaystyle\frac{g^{bc}}{\sqrt{-g}}
(\bar\theta\tilde\Pi_c{\Gamma^\mu}\epsilon)-
2i\partial_ax^\nu(\bar\theta\Gamma^\nu\Gamma^\mu\epsilon)-\cr
&&\qquad -2(\bar\theta\epsilon)(\bar\theta\Gamma^\mu\partial_a\theta),\\
&& \delta\bar\psi=i\epsilon^{ab}[\bar\epsilon\Gamma^\mu(\partial_a
\bar\theta\Gamma^\mu\partial_b\theta)-2\partial_a\bar\theta
(\partial_b\bar\theta\epsilon)],\cr
&& \delta\phi=-i\phi^2(\bar\psi\epsilon),\nonumber
\end{eqnarray}
where $\tilde\Lambda\equiv\Lambda_\mu\Gamma^\mu$, $\tilde\Pi_c=
{\Pi_c}^\mu\Gamma^\mu$. The action is invariant
up to total derivative terms. These transformations are the analog of Eq.
(27), since in the physical sector they are reduced to $\delta\theta_a=
\sqrt2 \epsilon'_a$, $\delta\bar\theta_{\dot a}=-\sqrt2
\bar\epsilon'_{\dot a}$, $\delta x^i=0$.

Global supersymmetries of the action (7), corresponding to Eq. (29) can
also be presented. To find them, let us consider the following ansatz:
\begin{equation}
\begin{array}{l} \delta\theta=\tilde\Lambda\tilde\Pi_c\epsilon^c, \qquad
\delta\phi=-i\phi^2(\bar\psi\tilde\Pi_c\epsilon^c),\\
\delta x^\mu=4i(\Lambda\Pi_c)(\bar\theta\Gamma^\mu\epsilon^c)+
2i(\bar\theta\tilde\Pi_c\epsilon^c)\Lambda^\mu,\end{array}
\end{equation}
where we denoted
\begin{equation}
\epsilon^a_\alpha\equiv P^{-ab}\epsilon_{\alpha\,b}, \quad
P^{-ab}=\frac 12\Big(\frac{g^{ab}}{\sqrt{-g}}-\varepsilon^{ab}\Big),
\quad (\Lambda\Pi_c)\equiv\Lambda^\mu{\Pi_c}^\mu.
\end{equation}
Variation of the GS part of the action (7) under these transformations
looks like
\begin{eqnarray}
\lefteqn{\delta S_{GS}=\varepsilon^{ab}[-8(\bar\theta\Gamma^\mu\epsilon^c)
(\partial_a\bar\theta\Gamma^\mu\partial_b\theta)(\Lambda\Pi_c)-
4(\bar\theta\tilde\Pi_c\epsilon^c)(\partial_a\bar\theta
\tilde\Lambda\partial_b\theta)+}\cr
&& +2(\partial_a\bar\theta\Gamma^\mu \tilde\Lambda\tilde\Pi_c\epsilon^c)
(\bar\theta\Gamma^\mu\partial_b \theta)+(\bar\theta\Gamma^\mu
\tilde\Lambda\tilde\Pi_c\epsilon^c)(\partial_a\bar\theta\Gamma^\mu
\partial_b\theta)]-\cr
&& -2iP^{-ba}[4(\bar\theta\tilde\Pi_c\epsilon^c)(\partial_a\Lambda
\Pi_b)+2(\partial_a\bar\theta\tilde\Lambda\epsilon^c)(\Pi_b\Pi_c)-
(\bar\theta\tilde\Lambda\partial_a\tilde\Pi_b\tilde\Pi_c
\epsilon^c)].
\end{eqnarray}
After integration by parts, reordering the $\tilde\Lambda$
and $\tilde\Pi$ terms and making use of the identities
\begin{equation}
\begin{array}{l}
P^{-ab}P^{-cd}=P^{-cb}P^{-ad},\\
(\partial_a\bar\theta\Gamma^\mu\partial_b\theta)(\Lambda\Pi_c)=
-\displaystyle\frac 12 \partial_a\bar\theta\Gamma^\mu\{\tilde\Lambda,
\tilde\Pi_c\}\partial_b\theta,\end{array}
\end{equation}
one can present all the terms in Eq. (33) as either
$K\tilde\Lambda\theta$ or $\partial_a\Lambda^\mu T^{\mu a}$, with
$K$ and $T$ some coefficients. These terms can evidently be cancelled by
appropriate variations of the $\bar\psi$ and $A_\mu^a$ variables in
the action. The final form for those variations is
\begin{eqnarray}
\lefteqn{\delta A^\mu_a=8(\bar\theta\Gamma^\rho\epsilon^c)
(\bar\theta\Gamma^\mu\Pi^\nu_c\Gamma^{\nu\rho}\partial_a\theta)-
5(\bar\theta\tilde\Pi_c\epsilon^c)(\bar\theta\Gamma^\mu
\partial_a\theta)-}\cr
&& -3\bar\theta\Gamma^\mu\Gamma^\nu\tilde\Pi_c\epsilon^c)
(\bar\theta\Gamma^\nu\partial_a\theta)-
4i\varepsilon_{ad}P^{-bd}[(\bar\theta\Gamma^\mu\epsilon^c)(\Pi_b\Pi_c)-\cr
&& -2(\bar\theta\tilde\Pi_c\epsilon^c)\Pi^\mu_b],\\[1ex]
\lefteqn{\delta\bar\psi=i\varepsilon^{ab}\{2(\partial_a\bar\theta
\tilde\Pi_c\epsilon^c)\partial_b\bar\theta-8(\partial_a\bar\theta
\tilde\Pi_c\partial_b\theta)\bar\epsilon^c-
8\partial_a[(\bar\theta\Gamma^\mu\epsilon^c)\partial_b\bar\theta
\Gamma^{\mu\nu}\Pi_\nu^c]+}\cr
&& +5(\bar\theta\partial_a\tilde\Pi_c\epsilon^c)\partial_b\theta
+3(\bar\theta\Gamma^\mu\partial_b\theta)\bar\epsilon^c\partial_a
\tilde\Pi_c\Gamma^\mu+(\partial_a\bar\theta\Gamma^\mu\partial_b\theta)
\bar\epsilon^c\tilde\Pi_c\Gamma^\mu\}-\cr
&& -2iP^{-ba}[\bar\epsilon^c\partial_a\tilde\Pi_c\tilde\Pi_b-
2\bar\epsilon^c(\Pi_b\partial_a\Pi_c)].\nonumber
\end{eqnarray}
Note that the complicated transformation low for the $\psi$-variable
can be predicted, since one of the Lagrangian equations of motion is
\begin{equation}
(\tilde\Lambda\psi)_\alpha=-4\tilde\Pi_b P^{-ba}\partial_a\theta_\alpha
+i\varepsilon^{ab}\theta^\beta\partial_a\theta^\gamma\partial_b
\theta^\delta\Gamma^\mu_{\alpha(\beta}(C\Gamma^\mu)_{\gamma\delta)}=0.
\end{equation}
Thus, transformation of the $\tilde\Lambda\psi$ part of the
$\psi$-variable is dictated by this equation and the transformation lows
for $x$ and $\theta$ variables.

Being reduced to the physical sector, Eq. (31) look as follows:
\begin{eqnarray}
&& \delta\theta_a=-\sqrt2 (P^+\epsilon_a-\partial_-x^i{\gamma^i}_{a\dot
a}\bar\epsilon'_{\dot a}+\partial_-x^{10}\epsilon'_a),\cr
&& \delta\bar\theta_{\dot a}=-\sqrt2 (P^+\bar\epsilon_{\dot a}
+\partial_-x^i\bar\gamma^i_{\dot aa}\epsilon'_a-
\partial_-x^{10}\bar\epsilon'_{\dot a}),\\
&& \delta x^i=2\sqrt 2 iP^+(\theta\gamma^i\bar\epsilon'-
\bar\theta\bar\gamma^i\epsilon')\nonumber
\end{eqnarray}
and seems to be an analog of Eqs. (29). Note that these transformations
act on the left moving modes only, in contrast to the eleven dimensional
superstring considered in Ref. 20. In this respect, the model presented
can be viewed as a $D=11$ analog of $N=1$, $D=10$ Green--Schwarz
superstring.

To summarize, in this letter we suggested a super Poincar\'e
invariant action for the closed superstring which classically exists
in any
spacetime dimension. From Eq. (26) it follows that zero modes of the
$\theta_a$, $\bar\theta_{\dot a}$ variables form the
Clifford algebra which is also symmetry algebra of a ground state.
A representation space is 256 dimensional and corresponds to the spectrum
of the $D=11$ supergravity [19]. Since supersymmetry is realized in the
physical subspace, one also gets the corresponding representation
in the space of functions on that superspace. This allows one to expect a
supersymmetric quantum states spectrum. Analysis of this situation in terms
of oscillator variables as well as the critical dimension will be
discussed in a separate publication.

The authors are grateful to I.L. Buchbinder for useful discussions.
The work of A.A.D. was supported by Joint DFG-RFBR project
No 96-02-00180G and project INTAS--96--0308. A.V.G. thanks ICTP for
the hospitality.


\begin{thebibliography}{nn}
\bibitem{1} M.B. Green and J.H. Schwarz, Phys. Lett. B {\bf 136} (1984)
367.
\bibitem{2} M.J. Duff, $M$-Theory (The Theory Formerly Known as
Strings), hep-th/9608117.
\bibitem{3} J.H. Schwarz, Lectures on Superstring and $M$-Theory
Dualities, hep-th/9607201.
\bibitem{4} M.J. Duff, Supermembranes, hep-th/9611203.
\bibitem{5} E. Witten, String Theory Dynamics in Various Dimensions,
hep-th/9503124.
\bibitem{6} E. Bergshoeff and E. Sezgin, and P.K. Townsend, Ann.
of Physics, {\bf 185} (1988) 330.
\bibitem{banks} T. Banks, W. Fischler, S. Shenker, and L. Susskind,
$M$-Theory as a Matrix Model: a Conjecture, hep-th/9610043.
\bibitem{7} L. Brink and M. Henneaux, {\it Principles of string
theory}, Plenum Press, New York and London, 1988.
\bibitem{8} T. Curtright, Phys. Rev. Lett. {\bf 60} (1987) 393.
\bibitem{9} E. Sezgin, Super $p$-Form Charges and Reformulation of the
Supermembrane Action in Eleven Dimensions, hep-th/9512082.
\bibitem{10} E. Bergshoeff and P.K. Townsend, Nucl. Phys. B
{\bf 490} (1997) 145.
\bibitem{pk} P.K. Townsend, $D$-Branes from $M$-Branes, hep-th/9512062.
\bibitem{11} M. Cederwall, A. von Gussich, B.E.W. Nilsson, and A.
Westenberg, The Dirichlet Super-Three-Brane in Ten-Dimensional Type IIB
Supergravity, hep-th/9610148.
\bibitem{12} M. Cederwall, A. von Gussich, B.E.W. Nilsson, P. Sindell,
and A. Westenberg, The Dirichlet Super-$p$-Brane in Ten-Dimensional
Type IIA and IIB Supergravity, hep-th/9611159.
\bibitem{13} M. Aganagic, C. Popesku, an J.H. Schwarz, Gauge-Invariant
and Gauge-Fixed $D$-brane Actions, hep-th/9612080.
\bibitem{14} A.A. Deriglazov and A.V. Galajinsky, Mod.Phys.Lett.A {\bf12}
(1997) 1517.
\bibitem{15} A.A. Deriglazov and A.V. Galajinsky, Phys. Lett. B {\bf
386} (1996) 141.
\bibitem{16} A.A. Deriglazov and A.V. Galajinsky, Phys. Rev. D {\bf 54}
(1996) 5195.
\bibitem{17} M. Kaku, {\it Introduction to Superstrings},
Springer-Verlag, 1988.
\bibitem{20} A.A. Deriglazov, Eleven dimensional superstring with new
supersymmetry and $D=10$ type IIA Green--Schwarz superstring,
hep-th/9709025.
\end{thebibliography}
\end{document}